\documentstyle[12pt]{article}

\begin{document}
\begin{titlepage}
\title{ Low-frequency instability of circularly polarized
microwave-pulsed-field-produced plasmas}
\author{B. Shokri\footnote {E-mail:B-shokri@sbu.ir}
\\{\small \it Physics Dept. and Laser Research Inst. of Shahid Beheshti
University, Evin,Tehran,Iran and}\\ {\small \it Inst. for studies
in Theoretical Physics and Mathematics, P.O.Box 19395-3351,
Tehran, Iran and}\\ {\small \it Dept. of Electrical and Computer
Engineering, University of Alberta, Edmonton, T6G 2V4, Canada}
\\M.~Ghorbanalilu\\ {\small \it Physics Department of Shahid
Beheshti University, Evin, Tehran, Iran}}
\end{titlepage}
\date{}
\maketitle
\begin{abstract}
The plasma produced by a high-power pulsed circularly polarized
microwave field which is  much weaker than the atomic fields is
investigated in the non-relativistic limit. It is shown that the
resulting electron distribution function is non-equilibrium and
anisotropic. Furthermore, the  dispersion equations derived to
study and analyze low-frequency ion oscillations. It is shown an
instability may grow  in the aforementioned system only when the
ion density exceeds a critical value.
\end{abstract}

\newpage
\vskip 1cm {\bf\large I. INTRODUCTION} \vskip 0.5cm
  A gas
   discharge in a microwave (MW) field is a rather complicated phenomenon that exhibits a
   variety of features. Numerous theoretical and experimental studies have been devoted to
   this phenomenon. Interest in such a  discharge is related not only to its potential for
   technological applications, but also  for general  understanding of the discharge
   physics.$^{1,2}$~A large amount  of experimental data  has been accumulated and
   many theoretical studies have been made to explain its various details. A number of
   experiments have been performed with radiation sources operating at wave lengths of~$8.5,
   4, 2$, and $0.8$~cm.$^3$ The strong MW field interaction with a gas raises interesting
   questions
    about the ionization mechanism and the electron distribution function
   created by this field. Furthermore, the interaction between high-power microwave fields
   with inhomogeneous plasmas results in a variety of interesting
    phenomena such as frequency upshift.$^{4}$

    In such a strong microwave wave field the
   electron oscillatory energy~ $\epsilon_{\textrm{\small  e}}$ is much higher than
   the ionization energy~$I_\textrm{{\small ioniz}}$ of the gas atoms
   \begin{equation}\label{1}
   \epsilon_\textrm{\small  e}=\frac{e^2E_0^2}{2m\omega_0^2}\gg I_\textrm{{\small ioniz}},
   \end{equation}
   where ~$\omega_0 $~is the radiation frequency, $m$ is the electron mass, and~ $E_0$ is
   the electric field strength. When the field amplitude
   is comparable to the atomic field strength~$E_\textrm{\small  a}\approx5.1\times
   10^9$~ V~cm$^{-1} $
   tunneling ionization becomes an important mechanism for direct ionization of gas atoms.
   This effect has been thoroughly studied in previous
   papers.$^{5-7}$~While condition~(1)~is satisfied by a
    large margin,
   relativistic effects are unimportant since the kinetic energy of electron oscillation in
   an electromagnetic field is comparable to the electron rest mass
   energy,  for  present day pulsed sources. In this case, gas atoms are primarily ionized by the
   impact of oscillating electrons
   and ionization is governed by the development of an
   electron avalanche.   Recently it has been shown   that
    discharge plasmas produced
    by  super strong pulsed MW fields are subject to the Weibel instability
    associated with the anisotropic nature of the electron distribution
    function.$^{8}$

    In the present paper, we consider the interaction of circularly polarized MW pulsed fields
    with frequencies ~$\omega_0\simeq 2\times10^{10}-2\times10^{11}~~$s$^{-1}$~~
    with a neutral gas in the non-relativistic regime; in particular we
     study low frequency ion oscillations and their
    instabilities.
 We will consider a pulsed radiation source which is capable of
    generating radiation  with an intensity of about $10^8~~$W~cm$^{-2}$,
    whose electric field~$ E_0 \leq 10^6~~ $V~cm$^{-1}$~is
    much weaker than the atomic field $E_\textrm{\small  a}$. In this case, we study
    the electron distribution  function (EDF) and the stability of
    the discharge plasma in the aforementioned frequency range at
    non-relativistic electron oscillation energy~$\epsilon_\textrm{\small  e}$~.

     This work is organized in five sections. In Sec. II we
    review the EDF generated by
   the interaction of linearly  polarized MW fields with a neutral
   gas. In Sec. III we   consider the effect of circularly MW fields
   on the plasma  produced and find the general
    dispersion equation for low frequency
    oscillations. In Sec. IV we solve the  dispersion equations for low frequencies to find an
    instability of the system  and its growth  rate. Finally, a summary and conclusions are presented in Sec. V.

   \vskip 1cm {\bf\large II. DISTRIBUTION FUNCTION }\vskip 0.5cm

Under condition~(1)~the thermal velocity of the electrons in a
discharge plasma can be neglected in
   comparison to the electron oscillation velocity in the MW radiation
   field. Since the collision frequency is much smaller than the MW
   field frequency, we can ignore
   the collisional randomization of the forced electron oscillation
    as well. Furthermore,
   if the plasma density~$n_\textrm{\small  e}(t)$  produced by the field during gas
   breakdown is less than the critical density
   (i.e.,~$\omega_0^2 > \omega_{\textrm{\small pe}}^2=4\pi ne^2 /m $) we can
   neglect the effect of the polarization field. Moreover, the plasma
   density is assumed to be much less
  than the neutral gas density~$n_0$~so that the latter can be
  considered constant.
  We also suppose that the field was adiabatically switched on
     in the infinite past. In addition, we can assume that the MW
     radiation electric
   field ~$\bf{E_0}$ is constant during a single field period.
  Therefore, by assuming that the electric field depends only on
  time,
   the kinetic equation for
  the plasma electrons produced in the gas breakdown by a
       strong pulsed field can be written as follows$^8$
     \begin{equation}\label{2}
\frac{\partial f_0}{\partial t}+e {\bf E_0}\cdot\frac{\partial
f_0}{\partial  {\bf p}}=n_0 \omega_\textrm{{\small ioniz}}\delta(
{\bf p}),
\end{equation}
      in the non-relativistic limit. Here $\delta({\bf p})$ is the
    Dirac delta function,
 $\omega_{\textrm{\small  ioniz}}$ is the ionization probability
of the gas atoms,
    and $n_0$ is the neutral   gas  number density.
  In the case of MW gas breakdown, electron-impact ionization is
  governed by the ionization probability $\omega_{\textrm{\small  ioniz}}$.
  From  Eq.   (2)  the
avalanche ionization parameter~$ \gamma(E_0)$ can be determined
.$^{7-9}$ We can neglect the right-hand side of Eq. (2) in the
first approximation and calculate the EDF directly by solving the
Vlasov equation under the following condition,$^{7-10}$
\begin{equation}\label{3}
  \omega_0\gg\gamma(E_0),\omega_{\textrm{\small ioniz}}.
\end{equation}
Condition (3) depends strongly on the neutral gas density and is
well satisfied at gas pressures of~$ p_0\simeq 10-100$ Torr. In
this approximation, to calculate the electron energy distribution
function, we assume the field components with circular
polarization to be
\begin{equation}\label{4}
  E_x=E_0 \sin \omega_0 t,
  ~~~~~~~~~~E_y=E_0 \cos \omega_0 t,
~~~~~~~~~~~~E_z=0,
\end{equation}
where the electric field amplitude $E_0$ describes a slowly
varying (over the field period) MW pulsed envelope.

Considering  condition (3) and solving Eq. (2), by  averaging the
distribution function over the MW field period, we find the
following expression for the EDF$^8$
\begin{equation}\label{5}
\left<\tilde{f_0}( v_\bot, v_z)\right>= \frac{\delta (v_z)}{2\pi^2
v_\bot \sqrt{4 v_E^2-v_\bot^2}},
\end{equation}
where, $f_0({\bf v},t)=n_e(t)\tilde{f_0}(\bf v)$ and
$v_E=eE_0/m\omega_0 $ is the electron oscillatory velocity in an
alternating electric field; $v_z$ is the perpendicular component
of electron velocity with respect to the  electric field; $v_\bot$
is the component of velocity in the plane of the electric field;
$\tilde{f_0}({\bf v})$ satisfies the normalization condition $\int
d{\bf v} \tilde{f_0}({\bf v})=1$.

 Since a pulse of a strong MW field actually contains on the
order of 10-100 wave periods the EDF is in principle unsteady.
However, as one would naturally expect the form of the EDF is
determined after several ionizating collisions. Thereafter the
electron concentration exponentially increases in time while the
shape of the EDF does not change. It is obvious that just on the
initial moment of the gas breakdown the  EDF is not in
equilibrium. But after several periods an EDF is produced such
that it usually characterizes the breakdown independent from
initial conditions. Furthermore,  the growth rate of the plasma
density is expressed by $\sim \gamma/\omega_0$. Thus, under the
gas low-pressure condition (3) the  velocity distribution function
of electrons is formed within the field period and is not
practically changed furthermore. In addition, during instability
development interested in the present paper the plasma density can
be considered constant. The rate of collective effects grows with
the increase of the plasma particles concentration, and this is
why the growth rates of plasma instabilities, usually small
compared to the plasma frequency, are proportional to the plasma
frequency.

Under conditions (1) and (3) the EDF is well described by Eq. (5)
since, at the  instant of originating, the electrons  are evenly
distributed in phase with the wave electric field.

\vskip 1cm {\bf\large III. DISPERSION EQUATION }\vskip 0.5cm

 The theory of the interaction of MW fields with a plasma has been widely
developed and covers an increasing range of plasma
phenomena.$^{11}$~In Sec. II the distribution function of the
charged particles for a homogeneous plasma produced by an external
circularly polarized MW electric field defined by Eq. (4) was
obtained. The stability of such a plasma can be analyzed by
considering  small perturbations of this distribution. To begin
with, we analyze MW fields with frequencies higher than all
characteristic plasma frequencies
\begin{equation}\label{7}
\omega_0 \gg \omega_{\textrm{\small p}\alpha},  \nu_\alpha~ ,
\end{equation}
where~$\omega_{\textrm{\small p}\alpha}$,~$ \nu_\alpha $ are the
plasma frequency,  and collision frequency for $\alpha$ species
(ions or electrons), respectively. To first-order approximation,
the plasma in a MW field may be regarded as isotropic.
Consequently, the field~${\bf\ E_0}(t)$ obeys the dispersion
equation for transverse wave
\begin{equation}\label{9}
k_0=\frac{\omega_0}{c}
\sqrt{1-\frac{\omega_{pe}^2}{\omega_0^2}}\simeq\frac{\omega_0}{c}.
\end{equation}
where $c$ is the speed of light in vacuum. Therefore, the MW field
can be considered homogeneous for processes with a characteristic
length of inhomogeneity $k^{-1}$ (where $k^{-1}$ is much smaller
than $k_0^{-1}$, or $k\gg\omega_0 /c$). Assuming this condition,
we analyze quasi-longitudinal oscillations in the stability of the
MW-field-produced plasma. In reality, the oscillation frequencies
are of the order of characteristic plasma frequencies, i.e.,
$\omega\sim \omega_{\textrm{\small pe}}$. Taking into account
relation (7), we find  the validity condition of the
quasi-longitudinal approximation to be $\omega\ll \omega_0 \simeq
k_0 c \ll k c $.

The MW electric field with circular polarization primarily
interacts with the neutral gas and produces a plasma  whose EDF
and properties were summarized in Sec. II. Since we will study
instabilities whose growth rates are higher than the ionization
rate we can assume a constant electron density. It is clear that,
for relatively weak fields in real plasmas, electrons are
accelerated to velocities significantly exceeding their thermal
velocity during the period of oscillatory electric field and the
mean free time. The result is a state in which electrons are
moving with respect to motionless ions. Such a charged particle
beam is unstable and its growth increment is rather large, of the
order of the plasma ion frequency or even larger. In this section
we derive the general dispersion relation for the MW produced
plasma in the limit of low-frequency ion oscillation
perturbations. Since the effect of a high frequency electric field
on the plasma ions can be neglected compared to its effect on the
electrons, the ion velocity distribution function can be regarded
as isotropic (usually Maxwellian) and the EDF is given by Eq. (5).

Linearizing the Vlasov equations for electrons and ions in a
collisionless plasma
we obtain the following equations
\begin{equation}\label{8}
\left.
\begin{array}{c}
\left.
\begin{array}{c}
\displaystyle \frac{\partial \delta f_\textrm{\small  e}}{\partial
t}+\imath{\bf k}\cdot{\bf v}~ \delta f_\textrm{\small  e}+\frac{e
E_0}{m}(\hat{{\bf i}}\sin\omega_0 t+\hat{{\bf j}}\cos\omega_0
t)\cdot \frac{\partial \delta f_\textrm{\small e}}{\partial {\bf
v}}+
\frac{e {\bf E}}{m}\cdot\frac{\partial f_{0\textrm{\small e}}({\bf
p}-{\bf p_0})}{\partial {\bf v}}=0
\end{array}
\right.
\end{array}
\right.
\end{equation}
\begin{equation}\label{9}
\displaystyle \frac{\partial\delta f_\textrm{\small i}}{\partial
t}+\imath{\bf k}\cdot{\bf v}~ \delta f_\textrm{\small i}+
\frac{e_\textrm{\small i} {\bf E}}{M}\cdot\frac{\partial
f_{0\textrm{\small i}}({\bf\ p})}{\partial {\bf\ v}}=0,
\end{equation}
for small deviations from the zero-order distributions. Here,
$\hat{{\bf i}}$ and $\hat{{\bf j}}$ are unit vectors along x and y
direction, respectively; ~$f_{0\textrm{\small e}}({\bf p}-{\bf
p_0})$~and~$f_{0\textrm{\small i}}({\bf p})$ are the zero-order
distribution functions of the electrons and the ions,
respectively; $e$ , $m$, $e_\textrm{\small i}$, and $M$ are the
electron and ion charge and mass, respectively and ${\bf p_0} = e
\int^t {\bf E_0} dt= -e {\bf E_0} \cos\omega_0 t /\omega_0$ and
$\imath=\sqrt{-1}$.  The small perturbations $\delta
f_\textrm{\small e}$~and~$\delta f_\textrm{\small i}$ depend on
the coordinates in the form of~$\exp(\imath{\bf k}.{\bf x})$,
because of the assumption of homogeneity;~${\bf\
E}=-\nabla\phi$~is the electric field due to the perturbations
satisfying the Poisson equation
\begin{equation}\label{10}
k^2 \phi = 4 \pi e\int d{\bf p}~ \delta f_\textrm{\small  e}+4 \pi
e_\textrm{\small i}\int d{\bf p}~ \delta f_\textrm{\small i}.
\end{equation}

In order to solve  Eqs. (8)-(10) it is convenient to introduce a
new  function
\begin{equation}\label{11}
\psi_\textrm{\small  e}({\bf\ p},t)=\exp\left\{-\frac{\imath
eE_0}{m \omega_0^2}(k_x\sin\omega_0 t+k_y\cos\omega_0 t)\right\}~
f_\textrm{\small  e} ({\bf p} +\displaystyle\frac{e
E_0}{\omega_0}[{\bf i}\cos\omega_0t-{\bf j}\sin\omega_0t]),
\end{equation}
where $k_x$ and $k_y$ are the component of the wave vector ${\bf k}$. The system of
equations then takes the form
\begin{equation}\label{12}
\left.
\begin{array}{c}
\left.
\begin{array}{c}
\hskip -5cm \displaystyle \frac{\partial \psi_\textrm{\small
e}}{\partial t}+ \imath{\bf k}\cdot{\bf v}~\psi_\textrm{\small  e}
-\imath{\bf k}\cdot \frac{\partial
f_{0\textrm{\small e}}({\bf p})}{\partial {\bf v}}~\frac{4\pi e}{m k^2} \\
\times\left[ e~\int d{\bf p}~ \psi_\textrm{\small  e}({\bf p})+
e_\textrm{\small i}~\int d{\bf p}~ \delta f_\textrm{\small i}({\bf
p})~ \exp\left\{-\displaystyle\frac{\imath eE_0}{m
\omega_0^2}(k_x\sin\omega_0 t+k_y\cos\omega_0 t)\right\}\right]=0,
\\[5mm]
\hskip -5cm \displaystyle \frac{\partial \delta f_\textrm{\small
i}}{\partial t}+\imath{\bf k}\cdot{\bf v}~\delta f_\textrm{\small
i}
-\imath{\bf k}\cdot \frac{\partial
f_{0\textrm{\small i}}({\bf p})}{\partial{\bf v}}~\frac{4\pi e_\textrm{\small i}}{M k^2}\\
\times\left[ e_\textrm{\small i}~\int d{\bf p}~ \delta
f_\textrm{\small i}+ e~\int d{\bf p}~\psi_\textrm{\small  e}~
\exp\left\{-\displaystyle\frac{\imath eE_0}{m
\omega_0^2}(k_x\sin\omega_0 t+k_y\cos\omega_0 t)\right\}
\right]=0.
\end{array}
\right.
\end{array}
\right.
\end{equation}
Applying the expansion
\begin{equation}\label{13}
\left.
\begin{array}{c}
\left.
\begin{array}{c}
\hskip -2.25cm \exp\left\{\pm \imath (k_x r_E \sin\omega_0 t+k_y
r_E\cos\omega_0 t)\right\}=
\\[5mm]
   \sum_{p=-\infty}^{+\infty}~\sum_{m=-\infty}^{+\infty}~
   \exp\left\{\pm \imath (p+m)\omega_0 t \right\}~\exp\left\{\pm \imath \displaystyle\frac{m\pi}{2}\right\}~J_p(k_x r_E)~J_m(k_y r_E),
\end{array}
\right.
\end{array}
\right.
\end{equation}
and introducing the decomposition
\begin{equation}\label{14}
\left(
\begin{array}{c}\psi_\textrm{\small  e}\\ \delta f_\textrm{\small
i}
\end{array}
\right)=\exp({-\imath\omega t})~\sum_{l=-\infty} ^{
+\infty}\sum_{n=-\infty} ^{+\infty}~ \exp\left\{\pm \imath
(l+n)\omega_0 t \right\}
\left(
\begin{array}{c}
\psi_{\textrm{\small e}_{ ln}}\\ \delta f_{\textrm{\small
i}_{ln}}\end{array} \right),
\end{equation}
we find from Eq. (12) the following equations
\begin{equation}\label{15}
\left.
\begin{array}{c}
\left.
\begin{array}{c}
\displaystyle -\imath\left[\omega + (l+n)
\omega_0\right]\psi_{\textrm{\small e}_{ln}}+\imath{\bf k}\cdot
{\bf v}~\psi_{\textrm{\small e}_{ln}}
-\imath{\bf k}\cdot\frac{\partial f_{0\textrm{\small e}}}{\partial
{\bf v}}~
\frac{4 \pi e} {m k^2} \\
\times \left[e~\int d{\bf p}~\psi_{\textrm{\small
e}_{ln}}+e_\textrm{\small i}~\int d{\bf p}~
\sum_{p=-\infty}^{+\infty}~ \sum_{m=-\infty}^{+\infty} \delta
f_{\textrm{\small i}_{pm}} \exp\left\{- \imath
\displaystyle\frac{(n-m)\pi}{2}\right\}~J_{l-p}(k_x
r_E)~J_{n-m}(k_y r_E)\right]=0~,
\\[5mm]
\displaystyle -\imath\left[\omega + (l+n) \omega_0\right] \delta
f_{\textrm{\small i}_{ln}}+ \imath{\bf k}\cdot {\bf v}~\delta
f_{\textrm{\small i}_{ln}}
-\imath{\bf k}\cdot\frac{\partial f_{0\textrm{\small i}}}{\partial
{\bf v}}~ \frac{4 \pi
e_\textrm{\small i}}{M k^2} \\
\times \left[e_\textrm{\small i}~\int~ d{\bf p}~\delta
f_{\textrm{\small i}_{ln}}+e~\int d{\bf p}~
\sum_{p=-\infty}^{+\infty}~ \sum_{m=-\infty}^{+\infty}
\psi_{\textrm{\small e}_{ln}} \exp\left\{\imath
\displaystyle\frac{(m-n)\pi}{2}\right\}~J_{p-l}(k_x
r_E)~J_{m-n}(k_y r_E) \right]=0,
\end{array}
\right.
\end{array}
\right.
\end{equation}
where~${\bf r}_E=e {\bf E_0}/m \omega_0 ^2$~is the amplitude of
the electron oscillation in the MW electric field and  $J_{n}$ is
the Bessel function of order $n$.

Now using the notation
$$u_{\textrm{\small e}_{ln}}=e~\int d{\bf p}~\psi_{\textrm{\small e}_{ln}}~,~~u_{\textrm{\small i}_{ln}}=
e_\textrm{\small i}~\int d{\bf p}~\delta f_{\textrm{\small i}_{l
n}}~,$$ we can write the formal solution of Eq. (15) in the form
of a coupled set of equation, as follows:
\begin{equation}\label{16}
\left.
\begin{array}{c}
\left.
\begin{array}{c}
u_{\textrm{\small e}_{ln}}=-\delta \varepsilon_\textrm{\small e}
\left(\omega+(l+n)\omega_0,{\bf
k}\right)\times\\
\left[u_{\textrm{\small e}_{ln}}+\sum_{p=-\infty}^{+\infty}~
\sum_{m=-\infty}^{+\infty}~ u_{\textrm{\small i}_{pm}}~
 \exp\left\{\imath
\displaystyle\frac{(m-n)\pi}{2}\right\}~ J_{p-l}(k_x
r_E)~J_{m-n}(k_y r_E)\right],
\\[5mm]
u_{\textrm{\small i}_{ln}}=-\delta \varepsilon_\textrm{\small i}
\left(\omega+(l+n)\omega_0,{\bf
k}\right)\times\\
\left[u_{\textrm{\small i}_{ln}}+\sum_{p=-\infty}^{+\infty}~
\sum_{m=-\infty}^{+\infty}~ u_{\textrm{\small e}_{pm}}~
 \exp\left\{-\imath
\displaystyle\frac{(n-m)\pi}{2}\right\}~ J_{l-p}(k_x
r_E)~J_{n-m}(k_y r_E)\right],
\end{array}
\right.
\end{array}
\right.
\end{equation}
where~$\delta \varepsilon_\textrm{\small  e}(\omega,{\bf\
k})$~and~$\delta \varepsilon_\textrm{\small i}(\omega,{\bf\ k})$~
are the partial contributions of the electron and ion
susceptibilities  to the longitudinal dielectric permittivity$^8$
$$
\varepsilon(\omega,{\bf k})=1+\delta\varepsilon_\textrm{\small
i}~+~\delta\varepsilon_\textrm{\small  e}.
$$
 The existence of solutions to this infinite set of coupled equations
gives the dispersion equation for small-amplitude longitudinal
oscillations in the external MW electric field. Since it is a
determinant of infinite order, it is impossible to analyze in
general. However, for the most interesting limiting cases this
determinant can be  simplified and the oscillation spectrum can be
found. For example, all the values~$\delta
\varepsilon_{\textrm{\small e}, \textrm{\small i}}(\omega+n
\omega_0,{\bf\ k})$~ with $n\neq0$, are small compared to unity in
the very high frequency limit, when condition (6) is satisfied.
Assuming this limit, only ~$u_{\textrm{\small
e}_{00}}$~and~$u_{\textrm{\small i}_{00}}$~are nonzero in Eq.
(16); thus, we obtain

\begin{equation}\label{17}
\left.
\begin{array}{c}
\left.
\begin{array}{c}
\left[1+\delta \varepsilon_\textrm{\small  e} (\omega,{\bf
k})\right] u_{\textrm{\small e}_{00}} + \delta
\varepsilon_\textrm{\small  e}(\omega,{\bf k})J_0(k_x r_E)~J_0(k_y
r_E)~u_{\textrm{\small i}_{00}}=0,
\\[5mm]
J_0(k_x r_E)~J_0(k_y r_E)~\delta \varepsilon_\textrm{\small
i}(\omega,{\bf k}) u_{\textrm{\small e}_{00}}+\left[1+\delta
\varepsilon_\textrm{\small i}(\omega,{\bf
k})\right]~u_{\textrm{\small i}_{00}}=0,
\end{array}
\right.
\end{array}
\right.
\end{equation}
 which gives the following dispersion equation
\begin{equation}\label{18}
1+\delta \varepsilon_\textrm{\small  e} (\omega,{\bf k})+\delta
\varepsilon_\textrm{\small  i}(\omega,{\bf k})+ \delta
\varepsilon_\textrm{\small  e}(\omega,{\bf k})~\delta
\varepsilon_\textrm{\small  i}(\omega,{\bf k})\left[1-J_0^2(k_x
r_E)~J_0^2(k_y r_E)\right]=0.
\end{equation}

By deriving Eq. (18), we fully neglected particle collisions.
However, this equation is also valid for collisional plasmas if
the oscillation velocity component of the particles is small
compared to the thermal velocity. The external MW field has no
effect on the collision process, i.e., on the scattering cross
sections.

\vskip 1cm {\bf\large IV. INSTABILITY OF THE MW PRODUCED
PLASMA.}\vskip0.5cm

The instability analyzed in this section occurs in the
range~$\omega_0\gg kv_E \approx \omega_{\textrm{\small pe}}$. It
only occurs for a  positive~(non-equilibrium)~ slope in the EDF
(5) and never
 for an equilibrium
 (Maxwellian)~electron energy
distribution  in this frequency range. Taking a Maxwell
distribution function for cold ions and the distribution function
(5) for electrons, the electron and ion susceptibilities
~$\delta\varepsilon_{\textrm{\small e}, \textrm{\small
i}}(\omega,{\bf\ k})$ are given by$^{8,11}$
\begin{equation}\label{19}
\left.
\begin{array}{c}
\left.
\begin{array}{c}
\hskip -6cm \displaystyle \delta \varepsilon_\textrm{\small
i}(\omega,k)=-\frac{\omega_{\textrm{\small pi}}^2}{\omega^2}~,
\\[5mm]
\displaystyle \delta \varepsilon_\textrm{\small e}(\omega,{\bf\
k})=\displaystyle \frac{\omega_{\textrm{\small pe}}^2}{k^2} \int
dv \displaystyle \frac{k
\partial \tilde{f_0}(v,t)/\partial
v}{\omega-kv}=-\displaystyle \frac{\omega_{\textrm{\small pe}}^2
k_\bot^2}{\omega^2 k^2}~\displaystyle
\frac{s}{\sqrt{1-\displaystyle \frac{4 k_\bot^2 v_E^2}{\omega^2}}}
 -\displaystyle \frac{\omega_{\textrm{\small pe}}^2 ~s}{\omega^2 \sqrt{(1-\displaystyle \frac{4
  k_\bot^2 v_E^2}{\omega^2})^3}}~,
\end{array}
\right.
\end{array}
\right.
\end{equation}
where
$$\displaystyle k_\bot=\sqrt{k_x^2+k_y^2}~,$$\vskip 0.2cm
\begin{equation}\label{20}
s=\left\{
\begin{array}{c}
\left.
\begin{array}{c}
 ~1~~~~~~~~~~ \mbox{for}~~~~~~~~~~~~~~~~~~~~~~~~ Re(\omega)\neq0\\
 1/2~~~~~~~\mbox{for}~~~~~~~~~~~~~~~~~~~~~~~~~~Re(\omega)=0
\end{array}
\right\}~~\mbox{when}~~~~~\displaystyle\frac{4k_\bot^2
v_E^2}{\omega^2}<1~,
\\[8mm]
\left.
\begin{array}{c}
 ~0~~~~~~~~~~ \mbox{for} ~~~~~~~~~~~~~~~~~~~~~~~~~ Im(\omega)=0\\
 ~1~~~~~~~~~~ \mbox{for}~ ~~~~~~~~~ Im(\omega)\neq0,Re(\omega)\neq0\\
 1/2~~~~~~~~~\mbox{for}~~~~~~~~~~~~~~~~~~~~~~~~~~Re(\omega)=0
\end{array}
\right\}~~\mbox{when}~~~~~\displaystyle\frac{4
k_\bot^2v_E^2}{\omega^2}>1~,
\end{array}
\right.
\end{equation}
where $Re(\omega)$ and $Im(\omega)$ are the real and imaginary
part of $\omega$, respectively. Substituting expressions~(19) and
(20)~into the dispersion equation~(18), we obtain the following
dispersion relations
\begin{equation}\label{21}
\left\{
\begin{array}{c}
\displaystyle\omega^2=\omega_{\textrm{\small
pi}}^2\left[1+\imath\left(\frac{\omega_{\textrm{\small pe}}^2
~\omega_{\textrm{\small pi}}}{|2k_\bot v_E|^3}-
\frac{\omega_{\textrm{\small pe}}^2
~k_\bot^2}{\omega_{\textrm{\small pi}}~|2k_\bot
v_E|~k^2}\right)J_0^2(\frac{k_\bot v_E}{\omega_0})\right]
~~~~~~~~~~~~~~~~~~~~4 k_\bot^2v_E^2\gg\omega_{\textrm{\small
pe}}^2~,
\\[5mm]
\displaystyle \omega=\frac{|2 k_\bot
v_E|~k_\bot}{k}\left[1-\frac{\imath}{2} \left(\frac{|2 k_\bot
v_E|^2~ k}{\omega_{\textrm{\small pe}}^2
~k_\bot}-\frac{\omega_{\textrm{\small pi}}^2
~k^3}{\omega_{\textrm{\small pe}}^2~k_\bot^3}\right)\right]
~~~~~~~~~~~~~~~~~~~~~~~~~~~~~4
k_\bot^2v_E^2\ll\omega_{\textrm{\small pe}}^2~,
\end{array}
\right.
\end{equation}
corresponding to unstable
low frequency ion oscillations. Here, the following approximation
\begin{equation}\label{22}
J_0^2(\frac{k_\bot v_E}{\omega_0})\approx J_0^2(\frac{k_x
v_E}{\omega_0})~J_0^2(\frac{k_y v_E}{\omega_0}),
\end{equation}
was used. Equation  (21) show  that in the case of a circularly
polarized MW field the low frequency oscillation takes place only
for perturbations satisfying the following condition:
\begin{equation}\label{23}
\frac{\omega_{\textrm{\small pi}}^2}{|2 k_\bot
v_E|^2}\geq\frac{k_\bot^2}{k^2}.
\end{equation}
This instability resulting from the positive slope of the
distribution function~(5)
can be interpreted as the stimulated Cherenkov excitation of the
low frequency ion oscillations by the radiation field of the
oscillating electrons. Perturbations that do not satisfy the
condition (23) are not unstable. Of course, according to
approximations and assumption used,  the instability satisfying
condition (23) can only occur when  the growth  rate exceeds the
ionization rate, i.e., the growth rate should exceed the avalanche
ionization constant~$\gamma(E_0)$.

\vskip 1cm {\bf\large V.
CONCLUSION } \vskip 0.5cm

In analyzing the interaction of a high frequency circularly
polarized MW field with a neutral gas, we obtained the general
dispersion relation for low frequency waves in the MW produced
plasma. From Eq. (23) for the instability condition it is
necessary that the
the ion density exceeds the critical value $n_i\geq M k_\bot^4
E_0^2/\pi m^2 k^2 \omega_0^2$. Under such condition  the electrons
transfer electric field energy to the ions and  stimulated
Cherenkov excitation of  low frequency ion oscillations takes
place. This effect requires a  positive slope in  the distribution
function and, in this case, the instability can propagate when the
grow rate
exceeds the avalanche ionization rate $\gamma(E_0)$. Other work
underway, is concerned with the case of linearly polarizd MW
fields.

\end{document}